\documentclass[aps,prd,twocolumn,showpacs,showkeys,nofootinbib]{revtex4}
\pdfoutput=1
\usepackage{epsfig}
\usepackage{amsmath}
\usepackage{array}
\usepackage{graphicx}
\newcommand{\beq}{\begin{equation}}
\newcommand{\eeq}{\end{equation}}
\newcommand{\bqa}{\begin{eqnarray}}
\newcommand{\eqa}{\end{eqnarray}}
\newcommand{\nnb}{\nonumber\\}

\newcommand{\bold}{\textbf}

\begin{document}

\title{Radiative $h_{c/b}$ decays to $\eta$ or $\eta^\prime$}

\author{Ruilin Zhu~\footnote{Email:rlzhu@njnu.edu.cn}}
\affiliation{
Department of Physics and Institute of Theoretical Physics, Nanjing Normal University, Nanjing, Jiangsu 210023, China}
\author{Jian-Ping Dai~\footnote{Email:daijianping@ihep.ac.cn}}
\affiliation{
INPAC, Shanghai Key Laboratory for Particle Physics and Cosmology, Department of Physics and Astronomy, Shanghai Jiao-Tong University,
Shanghai, 200240, China}
%\date{\today}

%%%%%%%%%%%%%%%%%%%%%%%%%%%%%%%%%%%%%%%%%%%%%%%%%%%%%%%%%%%%%%%%%%%%%%%%%%%%%%%%%%%%%%%%%%%%%%%%%%%%%%%%%%%%%%%%%%%%%%%%%%%%%%%%%
\begin{abstract}
Motivated by recent measurements of the radiative decay rates of the \emph{P}-wave spin singlet charmonium $h_c$ to the light meson $\eta$
or $\eta^\prime$ by the BESIII Collaboration, we investigate the decay rates of these channels at order $\alpha \alpha_s^4$. The
photon is radiated mainly from charm quark pairs in the lowest order Feynman diagrams, since the diagrams where a photon radiated
from light quarks are suppressed by $\alpha_s$ or the relative charm quark velocity $v$, due to Charge parity conservation. The form factors
of two gluons to $\eta$ or $\eta^\prime$ are employed, which are the major mechanism for $\eta$ and $\eta^\prime$ productions. $\eta(\eta^\prime)$
is treated as a light cone object when we consider that the parent charmonium mass is much heavier than that of the final light meson.
We obtain the branching ratio ${\cal B}(h_c\to \gamma\eta^\prime) = (1.94^{+0.70}_{-0.51})\times 10^{-3}$ in the nonrelativistic
QCD approach, which is in agreement with the BESIII measurement. The prediction of the branching ratio of $h_c\to \gamma\eta$ is
also within the range of experimental error after including the larger uncertainty of the total decay width $\Gamma_{h_c}$. The
applications of these formulae to the radiative decays to $\eta(\eta^\prime)$ of the \emph{P}-wave spin singlet bottomonium $h_b(nP)$ are
presented. These studies will shed some light on the $\eta - \eta^\prime$ mixing effects, the flavor SU(3) symmetry breaking, as
well as the nonperturbative dynamics of charmonium and bottomonium.
\end{abstract}
%%%%%%%%%%%%%%%%%%%%%%%%%%%%%%%%%%%%%%%%%%%%%%%%%%%%%%%%%%%%%%%%%%%%%%%%%%%%%%%%%%%%%%%%%%%%%%%%%%%%%%%%%%%%%%%%%%%%%%%%%%%%%%%%%

%\item[Keywords]
% 12.38.Bx   Perturbative calculations
% 13.25.Gv decays of jpsi,Upsilon and other quarkonia
% 14.40.Pq  heavy quarkonia

\maketitle
%%%%%%%%%%%%%%%%%%%%%%%%%%%%%%%%%%%%%%%%%%%%%%%%%%%%%%%%%%%%%%%%%%%%%%%%%%%%%%%%%%%%%%%%%%%%%%%%%%%%%%%%%%%%%%%%%%%%%%%%%%%%%%%%%
\section{Introduction}

Heavy quarkonium spectra have been systematically established over the past forty years. Over fifty mesons with the masses from 2.9~GeV
to 4.6~GeV, or from 9.3~GeV to 11.1~GeV, can be organized into charmonia and bottomonia, respectively~\cite{Brambilla:2010cs}. The
spin-parity quantum numbers $J^{PC}$ of these bound states or resonances can be understood through spin-orbital interactions and
radial excitations of a heavy quark-antiquark pair. Their masses can also be calculated in the framework of QCD potential model of
a heavy quark-antiquark pair~\cite{Gupta:1982kp,Ebert:2011jc}.

Along with the improvements of detecting precision and the accumulations of experimental data, many new charmonium-like and bottomonium-like
exotic states are discovered and some rare decay modes of conventional heavy quarkonia are observed~\cite{Agashe:2014kda}. The critical
issues on theoretical side are to determine the possible structures of the newly observed exotic states and precisely calculate the
decay rates for these heavy quarkonia and quarkonium-like states.

There are fewer studies on the rare decay modes of \emph{P}-wave spin singlet charmonium $h_c(n^1P_1)$ and bottomonium $h_b(n^1P_1)$. The
rare decay modes where both the heavy quark and antiquark are annihilated will provide more direct information on nonrelativistic
QCD (NRQCD) nonperturbative Long Distance Matrix Elements (LDMEs) of heavy quarkonia. Recently, based on a data sample of $4.5\times 10^8$
$\psi^\prime$ events, the decay branching fractions of $h_c\to \gamma\eta$ and $h_c\to \gamma\eta^\prime$ are measured respectively
to be $(4.7\pm1.5\pm1.4)\times 10^{-4}$ and $(1.52\pm0.27\pm0.29)\times 10^{-3}$ by the BESIII Collaboration~\cite{Ablikim:2016uoc},
where $h_c$ is produced in the decay $\psi^\prime\to \pi^0 h_c$.

The decay mode $h_c\to \gamma\eta(\eta^\prime)$ is interesting, since it is useful not only to extract the LDMEs of the P-wave spin
singlet charmonium but also to investigate the $\eta(\eta^\prime)$ production mechanism and the $\eta - \eta^\prime$ mixing effects.
According to flavor SU(3) symmetry, light mesons should be organized into two representations: singlet and octet. However, the flavor
SU(3) symmetry is a little broken, since the up ($u$), down ($d$), and strange ($s$) quarks have different masses. Thus $\eta$ and
$\eta^\prime$ may be viewed as the mixing states between flavor singlet and octet. More precisely, when considering the potential
gluonium content, $\eta$ can be treated as the mixing state between $|q\bar{q}\rangle$ and $|s\bar{s}\rangle$, while $\eta^\prime$
as the mixing state among $|q\bar{q}\rangle$, $|s\bar{s}\rangle$, and $|gg\rangle$~\cite{Muta:1999tc,Ali:2000ci}.

Considering the mass squared of emitted $\eta(\eta^\prime)$ is less than that of the parent heavy quarkonium, i.e. $m^2_{\eta(\prime)}<<m_{h_{c, b}}^2$,
a large momentum is transferred. $\eta(\eta^\prime)$ can be treated as a light cone object in the rest frame of parent heavy quarkonium.
Using the light cone approach, the form factors of two hard gluons transition to $\eta(\eta^\prime)$ can be employed.

In the paper, we shall establish the light-cone factorization for $\eta(\eta^\prime)$ production in the radiative $h_c$ decay, and
give the related phenomenological results. The factorization formulae can be applied to more channels: $h_c(n P)\to \gamma\eta(\eta^\prime)$,
$h_b(n P)\to \gamma\eta(\eta^\prime)$, and even $h_b(n P)\to \gamma\eta_c$ when letting $m_c<<m_b$.

This paper is organized as follows. In Sec.~\ref{formulae}, we present how to calculate the amplitudes. We briefly introduce the
NRQCD long-distance matrix elements and covariant projection method. Two gluons transition to $\eta^{(\prime)}$ form factors are
also given. The related phenomenological analyses are presented in Sec.~\ref{numerical}, where we shall give the numerical results
for the branching ratios of $h_c\to \gamma\eta(\eta^\prime)$, $h_b(n P)\to \gamma\eta(\eta^\prime)$, and $h_b(n P)\to \gamma\eta_c$.
In the end section we summarize the work and give a conclusion.

%%%%%%%%%%%%%%%%%%%%%%%%%%%%%%%%%%%%%%%%%%%%%%%%%%%%%%%%%%%%%%%%%%%%%%%%%%%%%%%%%%%%%%%%%%%%%%%%%%%%%%%%%%%%%%%%%%%%%%%%%%%%%%%%%
\section{Factorization formulae\label{formulae}}

\subsection{NRQCD LDMEs and covariant projection method}
The production and annihilation decays of heavy quarkonium have been systematically investigated more than twenty years, since a
rigorous NRQCD theory was established by Bodwin, Braaten, and Lepage~\cite{Bodwin:1994jh}. Though the factorization of hadroproduction
of heavy quarkonium has not proved up to date, the inclusive annihilation decays and many exclusive decays can be factorized into
all orders of Strong coupling constant $\alpha_s$~\cite{Bodwin:2008nf,Bodwin:2010fi}. The inclusive annihilation decay width of heavy
quarkonium $H$ can be factorized as~\cite{Bodwin:1994jh}
\begin{eqnarray}
  \Gamma(H) = \sum_n \frac{2\mathrm{Im} f_n(\mu_\Lambda)}{m_Q^{d_n - 4}} \langle H|{\cal O}_n(\mu_\Lambda)|H \rangle\, ,
\end{eqnarray}
where $\langle H|{\cal O}_n(\mu_\Lambda)|H \rangle$ are NRQCD LDMEs, which involve nonperturbative physics and are scaled by the
relative velocity $v$ between the heavy quark and antiquark with the mass $m_Q$ in the heavy quarkonium $H$. The short-distance coefficient $\mathrm{Im}f_n(\mu_\Lambda)$ can be calculated order by order in perturbative theory. The factor of $m_Q^{d_n - 4}$ has been introduced
so as to make the coefficient $f_n$ dimensionless.

\emph{P}-wave spin singlet charmonium $h_c$ exclusively decays to photon and $\eta(\eta^\prime)$ shall be factorized. In the following, one
can easily see there are no extra IR divergences at order $\alpha\alpha_s^{4}$. First, let us introduce the corresponding NRQCD LDMEs.
The related first two NRQCD operators which contributes to the above processes are
\begin{eqnarray}
  \mathcal{O}(^1P_1^{[1]}) &=& \psi^\dagger(-\frac{i}{2} {\overleftrightarrow{{\bold D}}}) \chi \cdot \chi^\dagger(-\frac{i}{2}
  {\overleftrightarrow{{\bold D}}}) \psi,\\
  \mathcal{O}(^1S_0^{[8]}) &=& \psi^\dagger T^a \chi\chi^\dagger T^a\psi,
\end{eqnarray}
where $\psi$ is the Pauli spinor field that annihilates a heavy quark while $\chi$ is the Pauli spinor field that crates a heavy
antiquark.

The matrix elements of the corresponding operators sandwiched by heavy quarkonium are usually denoted as
\begin{eqnarray}
  \langle \mathcal{O}(^{2S+1}L_J^{[1,8]}) \rangle_{H} &\equiv& \langle H|\mathcal{O}(^{2S+1}L_J^{[1,8]})| H\rangle.
\end{eqnarray}
The covariant projection method is useful to calculate the perturbative short-distance coefficients~\cite{Zhang:2005cha,Jia:2007hy,Zhu:2015jha}.
The Dirac spinors for the heavy quark with momentum $p_1$ and antiquark with momentum $p_2$ have the explicit forms
\begin{eqnarray}
  u_Q(p_1, \lambda) &=& \sqrt{\frac{E_1 + m_Q}{2E_1}}
  \left(
    \begin{array}{ll}
      ~~~~\xi_\lambda  \\
      \frac{\vec{\sigma} \cdot \overrightarrow{p_1}}{E_1 + m_Q} \xi_\lambda
    \end{array}
  \right)\,,
\end{eqnarray}
\begin{eqnarray}
  v_Q(p_2, \lambda) &=& \sqrt{\frac{E_2 + m_Q}{2E_2}}
  \left(
    \begin{array}{ll}
      \frac{\vec{\sigma} \cdot \overrightarrow{p_2}}{E_2 + m_Q} \xi_\lambda\\
      ~~~~\xi_\lambda
    \end{array}
  \right)\,,
\end{eqnarray}
where $E_1$ and $E_2$ are the energy of heavy quark and antiquark, respectively, which satisfy $E_1 = E_2 \equiv E$. $q$ is introduced
as half relative momentum between the heavy quark and antiquark with $p_H\cdot q = 0$, where $p_H = p_1 + p_2$. We have $E = \sqrt{m_Q^2 - q^2}$
in the heavy quarkonium rest frame. $\xi_\lambda$ is the corresponding two-component Pauli spinor and $\lambda$ is the polarization
quantum number. One can easily get the covariant expression for the spin-singlet and spin-triplet combinations of spinor bilinearities.
The projection operators can be written as
\begin{eqnarray}
  \Pi_S(q) &=& \sum_{\lambda_1, \lambda_2} u_Q(p_1, \lambda_1) \bar{v}_Q(p_2, \lambda_2) \langle\frac{1}{2}\lambda_1\frac{1}{2}\lambda_2|S S_z
  \rangle \nonumber  \\
           &=& -\frac{1}{4\sqrt{2}E(E + m_Q)} (\frac{1}{2} \, p\!\!\!\slash_{H} - q\!\!\!\slash + m_Q) \frac{p\!\!\!\slash_{H} + 2E}{2E}
  \nonumber  \\
           && \times\Gamma_S (\frac{1}{2} \, p\!\!\!\slash_{H} + q\!\!\!\slash - m_Q) \, ,
\end{eqnarray}
where $\Gamma_{S=1} = \varepsilon\!\!\!\slash(p_H) = \gamma^\mu \varepsilon_\mu(p_H)$ for the spin-triplet combination with the polarization
vector $\varepsilon_\mu(p_H)$, while $\Gamma_{S=0} = \gamma^5$ for the spin-singlet. The spin-triplet projection $\Pi_1(q)$ and the
spin-singlet projection $\Pi_0(q)$ are defined accordingly. Considering the color factor, one should add an extra factor $\bold{1}_c/\sqrt{N_c}$
where $\bold{1}_c$ is the unit matrix in the fundamental representation of the color SU(3) group.

To get the amplitudes for orbitally excited quarkonium involvements, one should do Taylor expansion for the amplitudes in powers
of half relative momentum $q^\mu$
\begin{eqnarray}
  {\cal A}(q) &=& {\cal A}(0) + \frac{\partial {\cal A}(q)}{\partial q^\mu}\mid_{q=0}q^\mu \nonumber  \\
  && +\frac{1}{2!}\frac{\partial^2 {\cal A}(q)}{\partial q^\mu\partial q^\nu}\mid_{q=0}q^\mu q^\nu + \ldots.
\end{eqnarray}
%

%%%%%%%%%%%%%%%%%%%%%%%%%%%%%%%%%%%%%%%%%%%%%%%%%%%%%%%%%%%%%%%%%%%%%%%%%%%%%%%%%%%%%%%%%%%%%%%%%%%%%%%%%%%%%%%%%%%%%%%%%%%%%%%%%
\subsection{$\eta - \eta^\prime$ mixing and two gluons transition form factors}

Because of  the flavor SU(3) symmetry breaking, the $\eta$ and $\eta^\prime$ mesons can not be well described by assigning them alone to
flavor octet and singlet. According to previous studies, there are two equally schemes to describe the $\eta - \eta^\prime$ mixing:
the flavor octet and singlet bases, and the quark-flavor bases~\cite{Akhoury:1987ed,Ball:1995zv,Kiselev:1992ms}. In the quark-flavor scheme, $\eta$ and $\eta^\prime$
can be treated as the mixing states between $|q\bar{q}\rangle$ and $|s\bar{s}\rangle$. But this is still not enough to explain $\eta^\prime$,
which is a really complicated and interesting meson. According to the current view, $\eta^\prime$ may have gluonium content, and
should be treated as the mixing state among $|q\bar{q}\rangle$, $|s\bar{s}\rangle$, and $|gg\rangle$~\cite{Ball:1995zv,Feldmann:1998vh,Escribano:2005qq,Frere:2015xxa,Muta:1999tc,Ali:2000ci}.

Defining the basis vectors $\eta_q = q\bar{q} = (u\bar{u} + d\bar{d})/\sqrt{2}$, $\eta_s = s\bar{s}$, the gluonium component
$\eta_g  = gg$, we have
\begin{eqnarray}\label{ds}
  |\eta \rangle        &=& \cos\phi \;|\eta_q\rangle - \sin\phi \;|\eta_s\rangle,  \\
  |\eta^\prime \rangle &=& \cos\phi_G(\sin\phi \;|\eta_q\rangle + \cos\phi \;|\eta_s\rangle) + \sin\phi_G \;|\eta_g\rangle, \nonumber  \\
\end{eqnarray}
where $\phi$ is the mixing angle between $|q\bar{q}\rangle$ and $|s\bar{s}\rangle$, while $\phi_G$ is introduced to include the gluonium
content of $\eta^\prime$.

Next we turn to the light cone distribution amplitudes for those components. The light cone distribution amplitudes of $\eta_q$ and
$\eta_s$ components in $\eta$ can be expanded in Gegenbauer Polynomial
\begin{equation}\label{gegen}
  \Phi^{(q, s)}_\eta(x, \mu) = 6x\bar{x}(1 + \sum_{n=1}^\infty a_{2n}(\mu) \;C_{2n}^{3/2}(x - \bar{x}))\;,
\end{equation}
where $x$ and $\bar{x} = 1 - x$ are the momentum fractions of the light quark and antiquark inside $\eta_{q, s}$, respectively. $B_n(\mu)$
can be evaluated through scale evolution at leading-order logarithmic accuracy
\begin{equation}\label{gegen1}
  a_{n}(\mu) = \left(\frac{\alpha_s(\mu)}{\alpha_s(\mu_0)}\right)^{\frac{\gamma_n}{\beta_0}} a_n(\mu_0)\;,
\end{equation}
where $\beta_0 = 11C_A/3 - 2n_f/3$ with flavor number $n_f$ and $\gamma_n$ reads as
\begin{equation}\label{gegen}
  \gamma_n = 4C_F(\psi(n+2) + \gamma_E - \frac{3}{4} - \frac{1}{2(n+1)(n+2)})\,,
\end{equation}
with the digamma function $\psi(n)$.

For $\eta^\prime$, the mixing effect between quark-antiquark and gluonium components should be taken into account. The corresponding
light cone distribution amplitudes are~\cite{Muta:1999tc,Ali:2000ci}
\begin{widetext}
\begin{eqnarray}
  && \Phi^{(q,s)}_{\eta^\prime}(x, \mu) = 6x\bar{x} \left\{1 + \left[a^{(q,s)}_2(\mu_0)
  \left(\frac{\alpha_s(\mu^2)}{\alpha_s(\mu_0^2)}\right)^{\frac{48}{81}} - \frac{a^{(g)}_2(\mu_0)}{90}
  \left(\frac{\alpha_s(\mu^2)}{\alpha_s(\mu_0^2)}\right)^{\frac{101}{81}}\right] C_2^{3/2}(x - \bar{x}) + \cdots\right\}\;,
  \nonumber \\
  && \Phi^{(g)}_{\eta^\prime}(x, \mu) = x\bar{x} (x - \bar{x}) \left[16a^{(q,s)}_2(\mu_0)
  \left(\frac{\alpha_s(\mu^2)}{\alpha_s(\mu_0^2)}\right)^{\frac{48}{81}} + 5a^{(g)}_2(\mu_0)
  \left(\frac{\alpha_s(\mu^2)}{\alpha_s(\mu_0^2)}\right)^{\frac{101}{81}}\right] + \cdots\;.
  \label{eq:gg}
\end{eqnarray}
\end{widetext}
Note that the above light cone distribution amplitude of gluonium component is different from the gluon light cone distribution amplitude
of Glueball~\cite{Zhu:2015qoa}, since the parity and charge parity are different.

\begin{figure}[th]
  \begin{center}
  \includegraphics[width=0.48\textwidth]{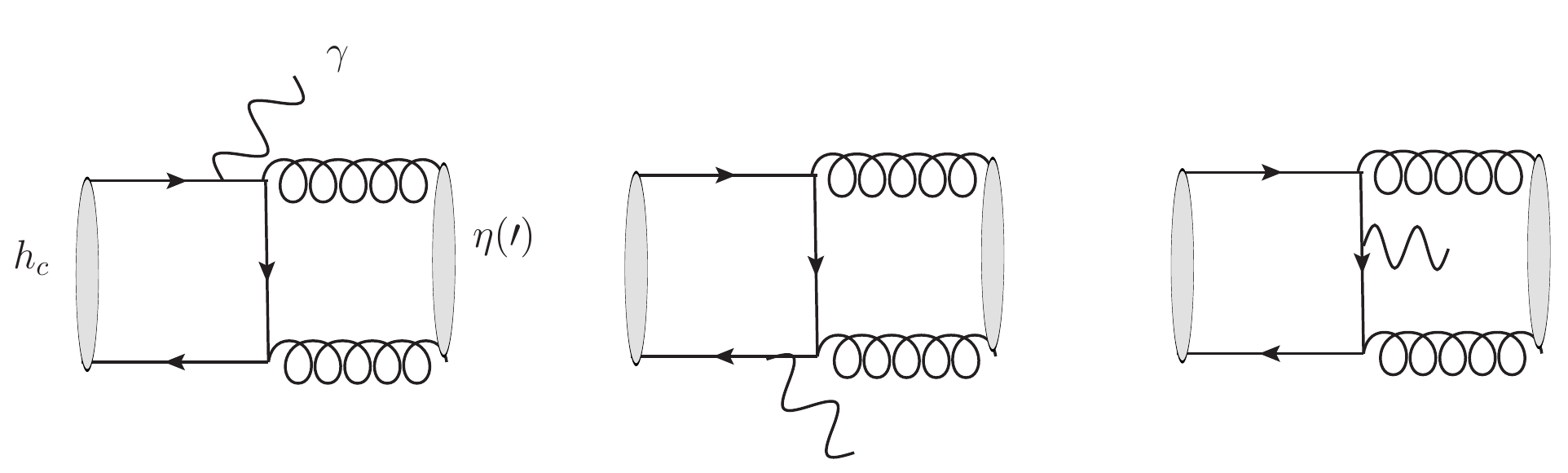}
  \end{center}
  \vskip -0.7cm
  \caption{Typical Feynman diagrams that contribute to the process $h_c\to \gamma\eta(\eta^\prime)$, where the effective vertex
  $g^*g^*\eta(\prime)$ denotes two gluons transitions to $\eta(\prime)$ form factors. Note that the diagrams where the photon is
  directly emitted from $\eta(\prime)$ and only two gluons link to $h_c$ contribute trivially, due to C-parity conservation.}
  \label{Fig-feynman}
\end{figure}

The studies on production mechanism for $\eta$ and $\eta^\prime$ are important, because they are useful to uncover the inner information
of $\eta$ and $\eta^\prime$ and to extract the mixing angles. Since $h_c$ has the spin-parity quantum numbers $1^{+-}$, it can not
directly decay to two gluons in order to preserve C-parity. The typical Feynman diagrams for the process $h_c\to \gamma\eta(\eta^\prime)$ are drawn in Fig.~\ref{Fig-feynman},
where the photon is emitted by charm quark pair.
Two gluons transition is the major mechanism for the production of $\eta$ and $\eta^\prime$ in radiative $h_c$ decays. This
kind of production mechanism is blind to quark charges, so the amplitude is identical to the production of $|q\bar{q}\rangle$ and $|s\bar{s}\rangle$.
 Two gluons transitions to $\eta$ and $\eta^\prime$ are also investigated in electroproduction
$e^+e^-\to J/\psi + \eta(\eta^\prime)$~\cite{Qiao:2014pfa}.

\begin{figure}[th]
  \begin{center}
  \includegraphics[width=0.42\textwidth]{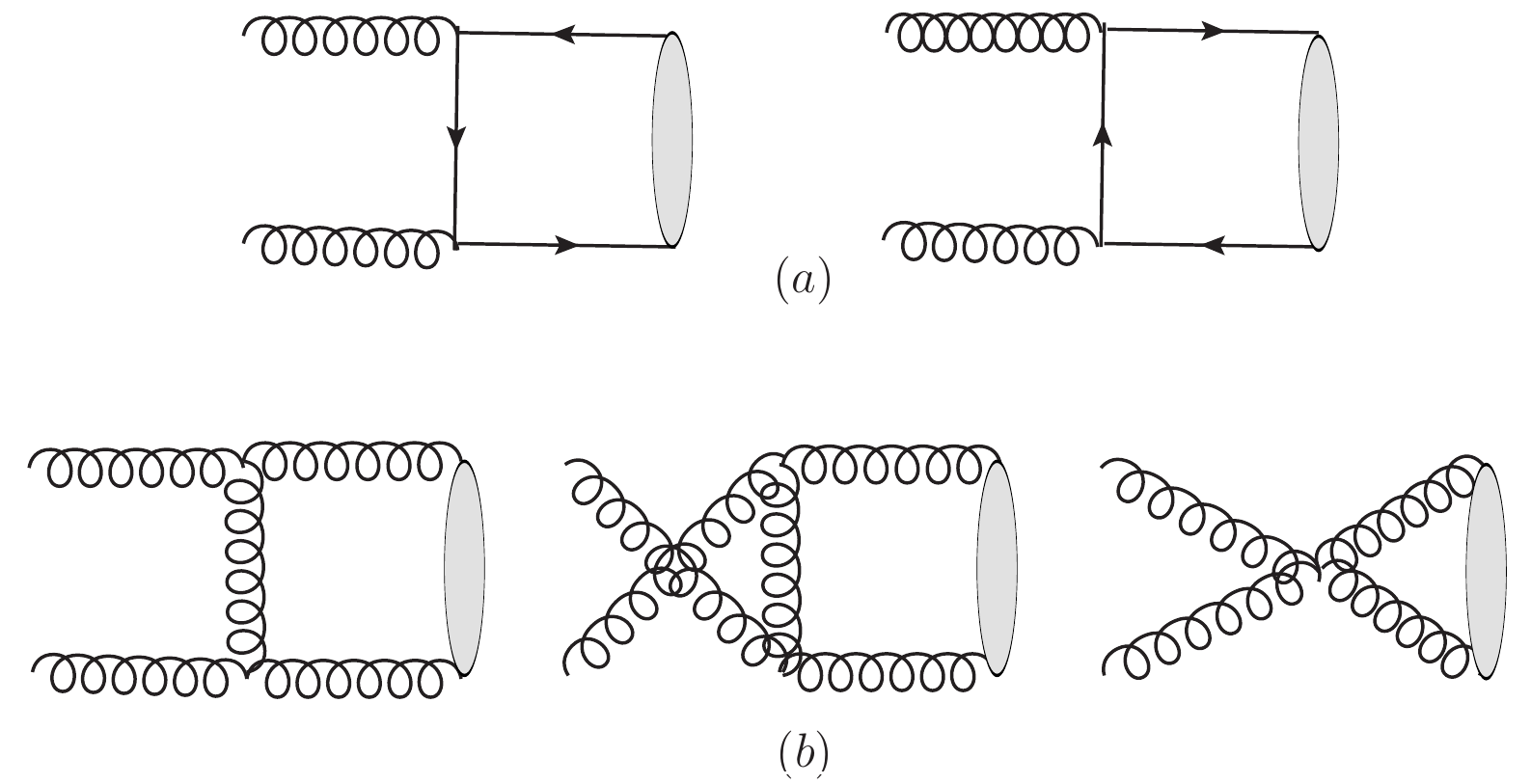}
  \end{center}
  \vskip -0.7cm
  \caption{Typical Feynman diagrams for two gluons transitions to $\eta^{(\prime)}$ form factors.}\label{Fig-formfactors}
\end{figure}

Two gluons transitions to $\eta^{(\prime)}$ form factors can be obtained by calculating the corresponding Feynman diagrams of two
gluons to $|q\bar{q}\rangle$, $|s\bar{s}\rangle$, and $|gg\rangle$, which are denoted in Fig.~\ref{Fig-formfactors}. The amplitudes
of two gluons to $|q\bar{q}\rangle$ and $|s\bar{s}\rangle$ are written as
\begin{equation}
  {\cal M}^{(q,s)} = -i\,F^{(q,s)}_{\eta^{(\prime)} g^*g^*}(q_1^2, q_2^2) \,\delta_{a b} \,\varepsilon^{\mu\nu\rho\sigma}
  \,\varepsilon^a_{1\mu} \varepsilon^b_{2\nu} q_{1\rho} q_{2\sigma},
  \label{eq:FFQ-1}
\end{equation}
where $q_1$ and $q_2$ are the momenta of two initial virtual gluons, respectively. $\varepsilon_1$ and $\varepsilon_2$ are the corresponding
polarization vectors of two initial gluons, respectively. Two gluons transitions to $\eta^{(\prime)}$ form factor $F^{(q,s)}_{\eta^{(\prime) g^*g^*}}$
is
\begin{eqnarray}
  && F^{(q,s)}_{\eta^{(\prime)} g^*g^*}(q_1^2, q_2^2) = \frac{2\pi\alpha_s(\mu^2)}{N_c} C_{\eta^{(\prime)}} \int_0^1 dx \Phi^{(q,s)}(x,\mu) \,
  \nonumber \\
  && \times \left[\frac{1}{xq_1^2 + \bar{x}q_2^2 - x\bar{x}m_{\eta^{(\prime)}}^2 + i\epsilon} + (x\leftrightarrow\bar{x})\right],
  \label{eq:QFF-2}\nnb
\end{eqnarray}
where $C_\eta = \sqrt{2} \, f^q_\eta + f^s_\eta$ and $C_{\eta^\prime} = \sqrt{2} \, f^q_{\eta^\prime} + f^s_{\eta^\prime}$. The decay
constants for the components are: $f^q_\eta = f_q \cos\phi$, $f^s_\eta = -f_s \sin\phi$, $f^q_{\eta^\prime} = f_q \sin\phi$ and
$f^s_{\eta^\prime} = f_s \cos\phi$.

The amplitude of two gluons to $|gg\rangle$ is expressed by
\begin{equation}
  {\cal M}^{(g)} = - i \, F^{(g)}_{\eta^\prime g^* g^*} \, \delta_{a b} \, \varepsilon^{\mu\nu\rho\sigma} \, \varepsilon^a_{1\mu}
  \varepsilon^b_{2\nu} q_{1\rho} q_{2\sigma} \,,
  \label{eq:FFG-def}
\end{equation}
and two gluons transitions to $\eta^{\prime}$ form factor $F^{(g)}_{\eta^{\prime} g^* g^*}$ is written as~\cite{Muta:1999tc,Ali:2000ci}
\begin{eqnarray}
  && F^{(g)}_{\eta^\prime g^* g^*}(q_1^2, q_2^2) = \frac{4\pi\alpha_s(\mu^2)}{Q^2} \, \frac{C_{\eta^\prime}}{2}
  \int_0^1 dx \, \Phi^{(g)}(x, \mu) \nonumber \\
  && ~~~~\times\left[\frac{x q_1^2 + \bar{x} q_2^2 - (1 + x\bar{x})m^2_{\eta^\prime}}{\bar{x} q_1^2 + x q_2^2 -
  x\bar{x}m^2_{\eta^\prime} + i\epsilon} - (x\leftrightarrow\bar{x})\right],
  \label{eq:GFF-result1}
\end{eqnarray}
where the parameter $Q^2$ is introduced to preserve the identical mass dimensions between the two transition form factors
$F^{{(q,s)}}_{\eta^{(\prime)} g^* g^*}$ and $F^{(g)}_{\eta^{\prime} g^* g^*}$. The choice of $Q^2$ has some freedom, which is usually
defined by the largest virtuality $q_1^2$ or $q_2^2$ ($|q_i^2|>>m_{\eta^\prime}^2$). In Ref.~\cite{Muta:1999tc,Ali:2000ci}, $Q^2$
is adopted to be $|q_1^2 + q_2^2|$. In order to calculate conveniently one-loop integrals, we adopt $Q^2 = q_1^2 + q_2^2 - m_{\eta^\prime}^2$,
where we assume hard scattering exists, i.e. $|q_i^2|>>m_{\eta^\prime}^2$. \\

%%%%%%%%%%%%%%%%%%%%%%%%%%%%%%%%%%%%%%%%%%%%%%%%%%%%%%%%%%%%%%%%%%%%%%%%%%%%%%%%%%%%%%%%%%%%%%%%%%%%%%%%%%%%%%%%%%%%%%%%%%%%%%%%%
\subsection{Decay amplitudes}

The amplitudes of $h_c\to \gamma\eta(\eta^\prime)$ can be parameterized into two independent terms
\begin{eqnarray}
  {\cal M}(h_c\to \gamma\eta(\eta^\prime)) &=& \frac{p_{h_c} \cdot \varepsilon(p_\gamma) \, p_\gamma \cdot \varepsilon^*(p_{h_c})}
  {p_\gamma \cdot p_{h_c}} F_-     \nonumber \\
  && +\varepsilon(p_{h_c}) \cdot \varepsilon^{*}(p_\gamma) F_+.
\end{eqnarray}
%
%%%%%%%%%%%%%%%%%%%%%%%%%%%%%%%%%%%%%%%%%%%%%%%%%%%%%%%%%%%%%%%%%%%%%%%%%%%%%%%%%%%%%%%%%%%%%%%%%%%%%%%%%%%%%%%%%%%%%%%%%%%%%%%%%
For the $^1P_1$ charmonium state, the summation over the polarization of $h_c$ is
\begin{eqnarray}
  \sum_{J_z = -1}^1 \varepsilon^\mu(p_{h_c}, J_z) \varepsilon^{*\nu}(p_{h_c},J_z) &=& -g^{\mu\nu} + \frac{p_{h_c}^\mu p_{h_c}^\nu}{m_{h_c}^2}.
\end{eqnarray}

The lowest order contribution can be written as
\begin{eqnarray}
  {\cal M} &=& \sum_{L_z}\langle 1L_z,00|1,J_z\rangle\langle 0|\chi^{\dagger}(-\frac{i}{2}{\overleftrightarrow{{\bold D}}})\psi|h_c\rangle
  \varepsilon^{\mu} (p_{h_c},L_z) \nonumber  \\
           && \times\varepsilon^{*\nu} (p_\gamma)\mathrm{Tr}[{\cal A}^{\mu\nu}(0)\Pi_0(0) + {\cal A}^{\nu}(0)\Pi^\mu_0(0)] ,
\end{eqnarray}
where
\begin{widetext}
\begin{eqnarray}
  {\cal A}^{\nu}(q) &=& \frac{4\pi e_c e \alpha_s C_A C_F}{ (m_{h_c} N_c)^{1/2}}\int \frac{d^4 k}{(2\pi)^4}
  \varepsilon^{\alpha\beta \rho \sigma} k_{\rho} (p_{\eta^{(')}}-k)_{\sigma} F^{(q, s, g)}_{\eta^{(\prime)} g^* g^*} (k^2, (p_{\eta^{(')}}-k)^2) \nonumber  \\
                    && \times\Big[ \frac{\gamma^\beta
  (m_c+p\!\!\!\slash_{\eta^{(')}}-k\!\!\!\slash-p\!\!\!\slash_2)\gamma^\alpha(m_c+p\!\!\!\slash_{\eta^{(')}}-p\!\!\!\slash_2)\gamma^\nu}
  {\left(\left(p_2-p_{\eta^{(')}}\right)^2-m_c^2\right)\left(\left(p_2+k-p_{\eta^{(')}}\right)^2-m_c^2\right)}+\frac{\gamma^\nu
  (m_c+p\!\!\!\slash_\gamma-p\!\!\!\slash_2)\gamma^\beta(m_c+p\!\!\!\slash_1-k\!\!\!\slash)\gamma^\alpha}
  {\left(\left(p_2-p_{\gamma}\right)^2-m_c^2\right)\left(\left(p_1-k\right)^2-m_c^2\right)} \nonumber  \\
                    && +\frac{\gamma^\beta
  (m_c+p\!\!\!\slash_{\eta^{(')}}-p\!\!\!\slash_2-k\!\!\!\slash)\gamma^\nu(m_c+p\!\!\!\slash_1-k\!\!\!\slash)\gamma^\alpha}
  {\left(\left(p_2+k-p_{\eta^{(')}}\right)^2-m_c^2\right)\left(\left(p_1-k\right)^2-m_c^2\right)}\Big],
  \label{Amu0}
\end{eqnarray}
\end{widetext}
with $p_1 = \frac{p_{h_c}}{2} - q$,  $p_2 = \frac{p_{h_c}}{2} + q$,
and
\begin{eqnarray}
  \Pi^\mu_0(0)= \frac{\partial {\Pi_0}(q)}{\partial q^\mu}\mid_{q=0}, \quad{\cal A}^{\mu\nu}(0) = \frac{\partial {\cal A}^{\nu}(q)}
  {\partial q^\mu}\mid_{q=0}.
\end{eqnarray}
Using the vacuum-saturation approximation for NRQCD LDMEs, $\langle H|\mathcal{O}_n| H\rangle\simeq \langle H|\psi^{\dagger}\mathcal{K}^\prime_n\chi|0 \rangle\langle 0|\chi^{\dagger}\mathcal{K}_n\psi| H\rangle$ with $\mathcal{O}_n = \psi^{\dagger}\mathcal{K}^\prime_n\chi\chi^{\dagger}\mathcal{K}_n\psi$.
Based on the power-counting rules~\cite{Bodwin:1994jh,Kramer:2001hh}, the matrix element $\mathcal{O}(^{2S+1}L_{J}^{[1,8]})_{H}$
scales as $v^{3+2L+2E+4M}$, where $S$, $L$, and $J$ are the spin and orbital angular momentum for $Q\bar{Q}$, the total angular
momentum for heavy quarkonium $H$. $E$ and $M$ are the minimum number of chromo-electric and chromo-magnetic transitions for $Q\bar{Q}$
from the lowest-order Fock state of $H$ to the state $Q\bar{Q}(^{2S+1}L_{J}^{[1,8]})$. The contribution of next-to-leading order Fock state
$^1 S^{[8]}_0$ will has a relative suppression $v^2_{c\bar{c}}\alpha_s$.

%%%%%%%%%%%%%%%%%%%%%%%%%%%%%%%%%%%%%%%%%%%%%%%%%%%%%%%%%%%%%%%%%%%%%%%%%%%%%%%%%%%%%%%%%%%%%%%%%%%%%%%%%%%%%%%%%%%%%%%%%%%%%%%%%
\section{Phenomenological discussions\label{numerical}}

The decay width of $h_c\to \gamma\eta(\eta^\prime)$ can be calculated using the following expression:
\begin{eqnarray}
  \Gamma(h_c\to \gamma\eta(\eta^\prime)) &=& \frac{|\textbf{p}|}{8\pi m_{h_c}^2}|{\cal M}|^2,
\end{eqnarray}
where
\begin{eqnarray}
  |\textbf{p}| &=& \frac{m_{h_c}^{2} - m_{\eta^{(')}}^2}{2m_{h_c}}, \nonumber
\end{eqnarray}
is the momentum modulus of $\eta(\eta^\prime)$ in the rest frame of $h_c$. The related branching ratio can be obtained by
${\cal B}(h_c\to \gamma\eta(\eta^\prime))= \Gamma(h_c\to \gamma\eta(\eta^\prime))/\Gamma_{h_c}$. The formulae can be employed to get
$h_b(n P)\to \gamma\eta(\eta^\prime)$, and $h_b(n P)\to \gamma\eta_c$ when assuming $m_b\gg m_c$.

In the numerical calculation we will adopt the parameters as follows: $m_{h_c} = 3.525$~GeV, $\Gamma_{h_c} = 0.7$~MeV, $m_{h_b} = 9.899$~GeV,
$m_{h_b(2P)} = 10.260$~GeV, $m_{\eta_c} = 2.984$~GeV, $m_{\eta} = 0.548$~GeV, $m_{\eta^{\prime}} = 0.958$~GeV~\cite{Agashe:2014kda}.
The heavy quark masses are $m_c = (1.5\pm0.5)$~GeV and $m_b = (4.8\pm0.5)$~GeV~\cite{Qiao:2012hp}. For the value of LDMEs for $h_c$, we adopt the result
of B-T potential model in Refs.~\cite{Buchmuller:1980su,Eichten:1995ch}
\begin{eqnarray}
  \langle\mathcal{O}(^1P_1^{[1]})\rangle_{h_c} = 0.1074(\mathrm{GeV})^5.
\end{eqnarray}

The decay constants for $\eta(\eta^\prime)$ are: $f_q = (1.07\pm0.02) f_\pi, \, f_s = (1.34\pm0.06)f_\pi$ with the pion decay constant
$f_\pi = 130.4$~MeV, and the mixing angle is adopted $\phi = 39.3^0\pm1.0^0$~\cite{Xiao:2014uza,Liu:2012ib}. The Gegenbauer momenta
are adopted as $a_2^{q,s}(1GeV) = 0.44\pm0.22$~\cite{Xiao:2014uza}. For $\eta^\prime$, we have $a_2^g(1GeV) = 0.1$ and $\mathrm{sin}^2 \phi_G = 0.26$~\cite{Qiao:2014pfa}. The decay constant of $\eta_c$ is about $498$~MeV~\cite{Qiao:2012hp}.

To manipulate the trace, the derivation of the decay amplitudes, and the matrix element squared, the Mathematica software is employed
with the help of the packages FeynCalc\cite{Mertig:1990an}, FeynArts\cite{Hahn:1998yk}, and LoopTools\cite{Hahn:2000jm}. The amplitudes
are Ultra-Violet and Infre-Red safe. The numerical results are listed in Tab.~\ref{tab:data}. The mixing among $|q\bar{q}\rangle$,
$|s\bar{s}\rangle$, and $|gg\rangle$ is important to explain the production properties of $\eta^\prime$. The result for the branching
ratio of $h_c\to \gamma\eta^\prime$ within NRQCD is consistent with the BESIII measurement~\cite{Ablikim:2016uoc}. While the prediction
of the branching ratio for $h_c\to \gamma\eta$ is smaller than the central value of the measurement~\cite{Ablikim:2016uoc}. Of course,
the uncertainty of the decay width of $h_c$ is large, i.e. $\Gamma_{h_c} = 0.7\pm0.4$~MeV~\cite{Agashe:2014kda}, which should be
considered and the branching ratio is changed accordingly. After including the uncertainty of the decay width $\Gamma_{h_c}$, the
branching ratios of $h_c\to \gamma\eta$ within NRQCD will be consistent with the BESIII data.

In Tab.~\ref{tab:data}, we also give the results for $h_b(n P)\to \gamma\eta_c$, which is reasonable when expanding the amplitudes
in order of $m_c/m_b$ and assuming $m_b\gg m_c$. The uncertainties for $h_c\to \gamma\eta(\eta^\prime)$ are large than other channels,
since we have set the scale at $m_c$ while the scale for other channels are set at $m_b$. When changing the heavy quark mass, the
strong coupling constant is also running.
\begin{table}[hthb]
  \caption{\label{tab:data} The branching rates of $h_c\to \gamma\eta(\eta^\prime)$, $h_b(n P)\to \gamma\eta(\eta^\prime)$, and
  $h_b(n P)\to \gamma\eta_c$. The uncertainties are from the heavy quark mass, and the scale is also set at the heavy quark mass.
  $\Gamma_{h_c} = 0.7$~MeV is adopted. }
  \begin{center}
  \setlength{\extrarowheight}{1.0ex}
  \renewcommand{\arraystretch}{1.3}
  \vspace{0.1cm}
  \begin{tabular}{cccccc}
  \hline\hline
    Branching rates                                               &This work               &Experiments~\cite{Agashe:2014kda}  \\
    \hline
    $10^{-4}{\cal B}(h_c\to \gamma\eta)$                          &$1.30^{+0.44}_{-0.32}$  &$4.7\pm1.5\pm1.4$     \\
    $10^{-3}{\cal B}(h_c\to \gamma\eta^\prime)$                   &$1.94^{+0.70}_{-0.51}$  &$1.52\pm0.27\pm0.29$  \\
    $\frac{{\Gamma}(h_b\to \gamma\eta)}{10^{-7}~MeV}$             &$2.64^{+0.24}_{-0.22}$  &-- \\
    $\frac{{\Gamma}(h_b\to \gamma\eta^\prime)}{10^{-6}~MeV}$      &$3.56^{+0.32}_{-0.30}$  &-- \\
    $\frac{{\Gamma}(h_b(2P)\to \gamma\eta)}{10^{-7}~MeV}$         &$3.20^{+0.27}_{-0.25}$  &-- \\
    $\frac{{\Gamma}(h_b(2P)\to \gamma\eta^\prime)}{10^{-6}~MeV}$  &$4.32^{+0.36}_{-0.33}$  &-- \\
    $\frac{{\Gamma}(h_b\to \gamma\eta_c)}{10^{-5}~MeV}$           &$1.13^{+0.12}_{-0.10}$  &-- \\
    $\frac{{\Gamma}(h_b(2P)\to \gamma\eta_c)}{10^{-5}~MeV}$       &$1.41^{+0.13}_{-0.12}$  &-- \\
    \hline\hline
  \end{tabular}
  \vspace{-0.2cm}
  \end{center}
\end{table}
%

%%%%%%%%%%%%%%%%%%%%%%%%%%%%%%%%%%%%%%%%%%%%%%%%%%%%%%%%%%%%%%%%%%%%%%%%%%%%%%%%%%%%%%%%%%%%%%%%%%%%%%%%%%%%%%%%%%%%%%%%%%%%%%%%%
\section{Conclusion}

In this paper, we have calculated the branching rates of the channels $h_c\to \gamma\eta(\eta^\prime)$ and $h_b(n P)\to \gamma\eta(\eta^\prime)$.
In these channels, two gluons transitions are the major production mechanism for $\eta(\prime)$ production. The amplitude of
$h_c\to \gamma\eta(\eta^\prime)$ is investigated, which is applied to the process $h_b(n P)\to \gamma\eta_c$ when assuming $m_b\gg m_c$.
The branching ratio of $h_c\to \gamma\eta^\prime$ within NRQCD is $(1.94^{+0.70}_{-0.51})\times 10^{-3}$, which is in agreement with
the BESIII measurement. After including the uncertainty of the total decay width $\Gamma_{h_c}$, the branching ratio of $h_c\to \gamma\eta$
within NRQCD is also consistent with the BESIII measurement. The BESIII detector will continue to collect the $\psi^\prime$ data for reaching
the total goal of $3\times 10^{9}$ events in future~\cite{Asner:2008nq}, which is nearly seven times as many as the sample used in
Ref.~\cite{Ablikim:2016uoc}. Thus the branching ratios of $h_c\to \gamma\eta(\eta^\prime)$ decay would be measured with higher precision.
In addition, the related $h_b$ decays could be explored in the future Belle II using high statistics data~\cite{Abe:2010gxa}. These
measurements will provide a unique method to study the $\eta - \eta^\prime$ mixing effects and the decay dynamics of charmonium and
bottomonium.

%%%%%%%%%%%%%%%%%%%%%%%%%%%%%%%%%%%%%%%%%%%%%%%%%%%%%%%%%%%%%%%%%%%%%%%%%%%%%%%%%%%%%%%%%%%%%%%%%%%%%%%%%%%%%%%%%%%%%%%%%%%%%%%%%
\section*{Acknowledgments}
This work was supported in part by the National Natural Science Foundation of China under Grant No. 11235005 and No. 11505111, and
the Research Start-up Funding of Nanjing Normal University, Natural Science Foundation of Shanghai under Grant No. 15DZ2272100 and
No. 15ZR1423100, by the Open Project Program of State Key Laboratory of Theoretical Physicsm, by the Open Project Program of State
Key Laboratory of Theoretical Physics, Institute of Theoretical Physics.

%%%%%%%%%%%%%%%%%%%%%%%%%%%%%%%%%%%%%%%%%%%%%%%%%%%%%%%%%%%%%%%%%%%%%%%%%%%%%%%%%%%%%%%%%%%%%%%%%%%%%%%%%%%%%%%%%%%%%%%%%%%%%%%%%

\end{document}